\documentstyle[aps,prl,twocolumn,epsfig]{revtex}

\preprint{hep-ph/0003169, CERN-TH/2000-077, DPNU-00-12, DUKE-TH-00-201}

\begin{document}

\title{Fluctuation Probes of Quark Deconfinement}
\author{Masayuki Asakawa$^1$, Ulrich Heinz$^2$, and Berndt M\"{u}ller$^3$\\
$^1$Department of Physics, Nagoya University, Nagoya 464-8602, Japan\\
$^2$Theoretical Physics Division, CERN, CH-1211 Geneva 23, Switzerland\\
$^3$Department of Physics, Duke University, Durham, NC 27708-0305}

\date{final version, \today}
\maketitle

\begin{abstract}
The size of the average fluctuations of net baryon number 
and electric charge in a finite volume of hadronic matter differs 
widely between the confined and deconfined phases. 
These differences may be exploited as indicators of the formation
of a quark-gluon plasma in relativistic heavy-ion collisions,
because fluctuations created in the initial state survive until 
freeze-out due to the rapid expansion of the hot fireball.
\end{abstract}


Fluctuations in the multiplicities and momen\-tum distributions of 
particles emitted in relativistic hea\-vy-ion collisions have been 
widely considered as probes of thermalization and the statistical 
nature of particle production in such reactions
\cite{Mrow,Shu98,BK99,NA49,BH99,DS99}. The characte\-ri\-stic behavior 
of temperature and pion multiplicity fluctuations in the final state 
has been proposed as a tool for the measurement of the specific heat 
\cite{Sto95} and, specifically, for the detection of a critical point 
in the nuclear matter phase diagram \cite{SRS}. Although the 
hot and dense matter created in heavy-ion collisions is not directly 
observed at the cri\-ti\-cal point (if one exists) but rather at the 
point of thermal freeze-out where particles decouple from the system, 
certain features of the critical fluctuations were shown to
survive due to the finite cooling rate of the fireball \cite{BR99}. 

We here draw attention to a different type of fluctuations which 
are sensitive to the microscopic structure of the dense matter. 
If the expansion is too fast for local fluctuations to follow the
mean thermodynamic evolution of the system, it makes sense to
consider fluctuations of locally conserved quantities that show
a distinctly different behavior in a hadron gas (HG) and a 
quark-gluon plasma (QGP). Characteristic features of the plasma 
phase may then survive in the finally observed fluctuations. This 
is most likely if subvolumes are considered which recede rapidly 
from each other due to a strong differential collective flow pattern 
as it is known to exist in the final stages of a relativistic 
heavy-ion reaction.

Three observables satisfy these constraints and are, in principle, 
measurable: the net baryon number, the net electric charge, and the 
net strangeness. Here we will focus on the first two as probes of 
the transition from hadronic matter to a deconfined QGP. Because 
they are sensitive to the microscopic structure of the matter, 
their unusual behavior would provide specific information about 
the structural change occurring as quarks are liberated and chiral 
symmetry is restored at high temperature. Our proposal differs from 
recent suggestions involving fluctuations in the abundance ratios of 
charged particles \cite{JK99} and in the baryon number multiplicity 
\cite{GP99} in that we only consider locally conserved quantities. 
We also disregard dynamical fluctuations of the baryon density caused 
by supercooling and bubble formation \cite{Gav99}.

We consider matter which is meson-dominated, i.e. whose baryonic 
chemical potential $\mu$ and temperature $T$ satisfy $\mu{\,\lesssim\,}T$. 
Our arguments will thus apply to heavy-ion collisions at CERN SPS energies 
and above. In the follo\-wing, we first explain qualitatively how 
hadronic and quark matter differ with respect to net baryon number 
and electric charge fluctuations. We then present analytical 
calculations supporting the argument. Finally, we estimate the 
rate at which initial state fluctuations are washed out during the 
expansion of the hot matter in the final, hadronic stage before 
thermal freeze-out.

In a hadron gas nearly two thirds of the hadrons (for $\mu\ll T$
mostly pions) carry electric charge $\pm 1$. In the deconfined
QGP phase the charged quarks and antiquarks make up only about half 
the degrees of freedom, with charges of only $\pm {1\over 3}$ or 
$\pm {2\over 3}$. Consequently, the fluctuation of one charged 
particle in or out of the considered subvolume produces a larger 
mean square fluctuation of the net electric charge if the system is 
in the HG phase. For baryon number fluctuations the situation is 
less obvious because in the HG baryon charge is now only carried by 
the heavy and less abundant baryons and antibaryons. Still, all of 
them carry unit baryon charge $\pm 1$ while the quarks and antiquarks 
in the QGP only have baryon number $\pm {1\over 3}$. It turns out 
that, as $\mu/T \to 0$, the fluctuations are again larger in the HG, 
albeit by a smaller margin than for charge fluctuations. At SPS 
energies and below the difference between the two phases increases 
since the stopped net baryons from the incoming nuclei contribute to 
the fluctuations, and more so in the HG than in the QGP phase.   
  
Generally, if $\cal O$ is conserved and $\mu$ is the associated 
chemical potential, in thermal equilibrium the mean square deviation 
of $\cal O$ is given by
 \begin{equation}
   (\Delta{\cal O})^2 \equiv 
     \langle {\cal O}^2 \rangle - \langle{\cal O}\rangle^2 =
     T\, {\partial\langle{\cal O}\rangle\over\partial\mu}\, ,
 \label{eq1}
 \end{equation}
where $\langle{\cal O}\rangle =  {\rm Tr\,} {\cal O}\, 
e^{-({\cal H}-\mu{\cal O})/T} \big/ {\rm Tr\,} 
e^{-({\cal H}-\mu{\cal O})/T}$. For ${\cal O}=N_b$ the r.h.s. 
of (\ref{eq1}) is $T$ times the {\em baryon number susceptibility} 
which was discussed earlier in the context of possible signatures 
for chiral symmetry restoration in the hadron-quark transition 
\cite{McL87}.

In general, the relative fluctuation of any extensive variable
vanishes in the thermodynamic limit $V{\,\to\,}\infty$ because the 
expectation value $\langle{\cal O}\rangle$ increases linearly with 
the volume $V$ while the fluctuation $\Delta{\cal O}$ grows only 
like $\sqrt{V}$. In reality, the value of a conserved quantum number 
of an isolated system does not fluctuate at all. However, if we 
consider a small part of the system, which is large enough to neglect 
quantum fluctuations, but small enough that the entire system can be 
treated as a heat bath, Eq.\,(\ref{eq1}) can be used to calculate 
the statistical uncertainty of the value of the observable in the 
subsystem. This is the scenario considered here.

We first discuss the fluctuations of the net baryon number. Since 
baryons are heavy, in the dilute HG phase we can apply the 
Boltzmann approximation \cite{fn1}:
 \begin{equation}
    N_b^{\pm}(T,\mu) = N_b^{\pm}(T,0)\, \exp(\pm\mu/T) \, .
 \label{eq3}
 \end{equation}
Here $N_b^{\pm}$ denotes the number of baryons $(+)$ and antibaryons
$(-)$, respectively. The net baryon number is 
$N_b\,{=}\,N_b^{+}{-}N_b^{-}$. Then the net baryon number 
fluctuations in the hadronic gas are given by
 \begin{equation}
    (\Delta N_b)_{\rm HG}^2 = N_b^+ + N_b^-
    = 2\, N_b^{\pm}(T,0)\, \cosh(\mu/T)\, . 
 \label{eq4}
 \end{equation}
This result makes sense, because the fluctuation of either a
baryon or an antibaryon into or out of the subvolume changes the
net baryon number contained in it.

To estimate $(\Delta N_b)^2$ in the QGP phase, we use the exact 
result for the baryon number density in an ideal gas of massless 
quarks and gluons (for two massless flavors):
 \begin{equation}
   {1\over V}\, (\Delta N_b)_{\rm QGP}^2 = {2\over 9} T^3
   \left(1 + {1\over 3} \Bigl({\mu\over \pi T}\Bigr)^2 \right) \,  ,
 \label{eq5}
 \end{equation}
where $V$ denotes the volume of the considered subsystem. It is
convenient to normalize this by the entropy density (again for two
quark flavors plus gluons):
 \begin{equation}
   {1\over V}\, S_{\rm QGP} = {74\pi^2\over 45}\, T^3 
   \left( 1 + {5\over 37} \Bigl({\mu\over\pi T}\Bigr)^2\right) \, .
 \label{eq6}
 \end{equation}
The later expansion being nearly isentropic, the ratio
 \begin{equation}
   \left. {(\Delta N_b)^2 \over S}\right\vert_{\rm QGP} 
   = {5\over 37\pi^2}
   \left( 1 + {22\over 111}\Bigl({\mu\over \pi T}\Bigr)^2 + \ldots
   \right) \, ,
 \label{eq7}
 \end{equation}
provides a useful measure for the fluctuations predicted for a
transient quark phase. The entropy can be estima\-ted from the final 
hadron multiplicity \cite{SH92}. 

For high collision energies ($\mu/T{\,\to\,}0$), the ratio (\ref{eq7})
ap\-proaches a constant; even for SPS energies, the $\mu$-de\-pen\-dent 
correction is at most 5\%. The many resonance contributions make it 
difficult to write down an analytic expression like (\ref{eq6}) 
for the entropy density in a hadron gas, but it is clear that the  
stronger $\mu$-dependence of (\ref{eq4}) compared to (\ref{eq5}) 
induces a stronger $\mu$-dependence of the corresponding ratio 
(\ref{eq7}) in the HG phase. This translates into a stronger beam 
energy dependence of the ratio (\ref{eq7}) near midrapidity in the 
HG than in the QGP phase.

Before providing numerical illustrations, let us compare these results
with those for net charge fluctuations. All stable charged hadrons have
unit electric charge; again using the Boltzmann approximation, which
only for pions introduces a small error of at most 10\%, we find 
 \begin{equation}
   (\Delta Q)_{\rm HG}^2 = N_{\rm ch} ,
 \label{eq8}
 \end{equation}
where $N_{\rm ch}$ is the total number of charged particles emitted
from the subvolume. To find the expression for a noninteracting QGP,
we introduce the electrochemical potential $\phi$ which couples to the
electric charges $q_u={2\over 3}$ and $q_d=-{1\over 3}$ of the up- and
down-quarks: 
 \begin{equation}
   {\langle Q(\phi) \rangle\over V} = \sum_{f=u,d} q_f 
   \Bigl(({\textstyle{1\over 3}}\mu{+}q_f\phi) T^2
         +{1\over\pi^2}({\textstyle{1\over 3}}\mu{+}q_f\phi)^3
   \Bigr) .
 \label{eq9}
 \end{equation}
We differentiate with respect to $\phi$ at $\phi=0$ and normalize to 
the entropy density:
 \begin{equation}
    \left. {(\Delta Q)^2 \over S}\right\vert_{\rm QGP} = {25\over 74\pi^2}
    \left( 1 + {22\over 111}\Bigl({\mu\over\pi T}\Bigr)^2 + \ldots
    \right) .
 \label{eq12}
 \end{equation}
This is a factor ${5\over 2}$ larger than the corresponding ratio
(\ref{eq7}) for baryon number fluctuations, due to the larger electric 
charge of the up-quarks, but shows the same weak $\mu$-dependence.
The main difference to baryon number fluctuations arises in the HG
phase: Since at SPS and higher energies the r.h.s. of (\ref{eq8}) is
dominated by pions and meson resonances, its $\mu$-dependence is now
also weak. In contrast to baryon number fluctuations, charge
fluctuations thus show a weak beam energy dependence in either phase, 
and only their absolute values differ \cite{fn2}.

We now give some numerical values for the fluctuation/entropy ratios
at SPS and RHIC/LHC. At the SPS, the net baryon number per unit of 
rapidity is measured: $dN_b/dy\approx 92$ \cite{NA49-97}. The 
antibaryon/baryon ratio is $\approx 0.085$ \cite{PBM99,NA49-96}, 
corresponding to $dN_b^{-}/dy\approx 8.5$. Combined with a specific 
entropy of $S/N_b\approx 36$ \cite{CR99}, Eq.~(\ref{eq4}) thus gives
$(\Delta N_b)^2/S \approx 0.033$ if the fluctuations reflect an equilibrium 
HG. If they have a QGP origin, Eq.~(\ref{eq7}) gives 
$(\Delta N_b)^2/S \approx 0.014$ \cite{fn1}, i.e.\,about a factor 
2.4 less. --- The charge fluctuations in a HG can be evaluated from 
the measured charged multiplicity density at midrapidity, 
$dN_{\rm ch}/dy \approx 400$ \cite{NA49-97,NA49-96,NA49-99}, after 
correcting for resonance decays \cite{SH92}. Assuming hadrochemical
freeze-out at $T\approx 170$\,MeV \cite{PBM99}, 60\% of the observed
pions stem from such decays \cite{SKH91}. One finds $(\Delta Q)^2/S
\approx 0.06$. If the charge fluctuations arise from a QGP,
Eq.~(\ref{eq12}) gives $(\Delta Q)^2/S \approx 0.036$, i.e. 60\% of
the HG value. 

\begin{figure}
\centerline{\epsfig{file=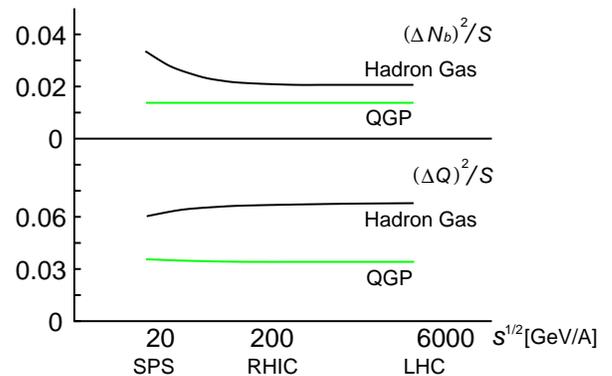,width=0.9\linewidth}}
\vskip 3mm
\caption{
  Schematic drawing of the beam energy dependence of the net
  baryon number and charge fluctuations per unit entropy
  for a hadronic gas and a quark-gluon plasma.}
\label{fig1}
\end{figure}

It is instructive to extrapolate these results to RHIC/ LHC energies
(i.e. $\mu/T{\,\to\,}0$). We again assume hadrochemical freeze-out 
at $T{\,\approx\,}170$\,MeV and use the particle multiplicities 
predicted by hadrochemical models \cite{Zim99,Sta99}. One obtains 
$(\Delta N_b)^2/S{\,\approx\,}0.020$ in the HG, compared to 0.0137 
in the QGP, and $(\Delta Q)^2/S \approx 0.067$ in the HG phase, 
compared to 0.034 in the QGP. Only the first of these four numbers, 
corresponding to the hadronic baryon number fluctuations, changes 
by more than 10\% as one proceeds from SPS to RHIC (see Fig.~\ref{fig1}). 

These estimates, including our corrections for resonance decays,
refer to ideal gases in equilibrium. Future work should address 
interaction effects on the thermal fluctuations in HG and QGP and 
treat resonance decays kinetically. We also point out potentially 
important non-equilibrium aspects: The fluctuation/entropy ratios in 
the QGP will be even lower (facilitating the discrimination against 
HG) if initially the QGP is strongly gluon-dominated \cite{Sh92} and 
hadronizes before the concentrations of the (baryon) charge carriers 
$q,\bar q$ saturate \cite{Betal}, or if hadronization itself 
generates additional entropy.  

We now discuss whether the difference between the two phases (typically 
a factor 2) is really observable. Even if a QGP is temporarily created 
in a heavy-ion collision, all hadrons are emitted after re-hadronization. 
Thus, it is natural to ask whether the fluctuations will not always 
reflect the hadronic nature of the emitting environment. We must show 
that the time scale for the dissipation of an initial state fluctuation 
is larger than the duration from hadronization to final particle 
freeze-out. It is essential to our argument that fluctuations of 
conserved quantum numbers can only be changed by particle transport 
and thus are likely to be frozen in at an early stage, similar to 
the abundances of strange hadrons, which are frozen early in the 
reaction and may even reflect the chemical composition of a 
deconfined plasma \cite{Bia98}.

For our estimate we assume for simplicity that the fire\-ball expands 
mostly longitudinally, with a boost-in\-va\-r\-ant (Bjorken) flow 
profile. Longitudinal position and rapidity are then directly related.
Strong longitudinal flow exists in collisions at the SPS \cite{NA49-HBT}, 
and the Bjorken picture is widely expected to hold for collisions 
at RHIC and LHC. Consider a slice of matter spanning a rapidity 
interval $\Delta\eta$ at the initial time $\tau_i$. ($\tau$ is the 
proper time and $\eta{\,=\,}\tanh^{-1}(z/t)$.) Its proper volume is 
$V_i{\,=\,}A\tau_i\Delta\eta$ where $A$ is the transverse area of the 
fireball. We denote the initial total baryon density by 
$\rho_i{\,=\,}\rho_{b{+}\bar b}(\tau_i)$. We assume
$T_i{\,=\,}170$\,MeV and $T_f{\,=\,}120$\,MeV for the initial and
final temperature \cite{Stock}, corresponding to
$\tau_i{\,\approx\,}2.5$\,fm/$c$ and $\tau_f{\,\approx\,}7$\,fm/$c$ at
the SPS, and $\tau_i{\,\approx\,}5$\,fm/$c$ and
$\tau_f{\,\approx\,}14$\,fm/$c$ at RHIC. 

Let us first give a qualitative argument \cite{KR} for the survival 
of a baryon number fluctuation within a rapi\-di\-ty interval 
$\Delta\eta{\,\approx\,}1$. Between $\tau_i$ and $\tau_f$, this 
interval expands from a length of 5\,fm to 14\,fm (we use the
RHIC numbers here). Baryons have average thermal longitudinal 
velocity component ${\bar v}_z{=}{1\over 2}{\bar v}$ where 
${\bar v}{\,\equiv\,}\langle|\bbox{v}|\rangle{\,=\,}\sqrt{8T/\pi M}$
is the mean thermal velocity ($\bar v{\,=\,}0.65$ for baryons 
with $M{\,=\,}1$\,GeV at $T{\,=\,}170$ MeV). Without rescattering, 
between $\tau_i$ and $\tau_f$ a baryon which is initially at the 
center of this interval can travel on average only about 3\,fm in 
the beam direction; hence it will not reach the edge of the interval 
before freeze-out. Because of rescattering in the hot hadronic matter, 
the baryon number actually diffuses more slowly, and a fluctuation 
will even survive in a smaller rapidity interval.

For a quantitative argument, we need to estimate the flux of baryons
in and out of the considered rapidity interval. Two effects need to be 
evaluated in this context. First, the difference in the baryon 
densities inside and outside the subvolume causes a difference in the
values of the mean flux of baryons into and out of the volume. Denoting
by ${\bar v}(\tau)$ the average thermal velocity of baryons, one finds
that the initial fluctuation decays exponentially:
 \begin{equation}
   \Delta N_b(\tau) = \Delta N_b^{(i)} \exp\left(
   -{1\over 2\Delta\eta} \int_{\tau_i}^{\tau} {d\tau\over\tau}\,
   {\bar v}(\tau) \right) \, .
 \label{eq13}
 \end{equation}
In the Bjorken scenario, the temperature $T$ falls as $\tau^{-1/3}$
and one finds for the remaining fluctuation at freeze-out
 \begin{equation}
   \Delta N_b(\tau_f) = \Delta N_b^{(i)} 
   \exp\left(-{3{\bar v}_i\over\Delta\eta}\, [1-(T_f/T_i)^{1/2}]\right)\, .
 \label{eq14}
 \end{equation}
For the numbers considered here, the exponent is very close to
$-{\bar v}_i/(2\Delta\eta)$, implying that the fluctuation survives
if $\Delta\eta$ is larger than ${\bar v}_i/2{\,\approx\,}0.33$.

The second effect that can wash out the initial fluctuation is
fluctuations in the baryon fluxes exchanged with the neighboring
subvolumes. These could eventually replace the initial fluctuation
with a thermal fluctuation that is characteristic of the conditions
at freeze-out. The total number of baryons entering $N_b^{(\rm en)}$
or leaving $N_b^{(\rm lv)}$ the subvolume between
$\tau_i$ and $\tau_f$ is given by
 \begin{equation}
   N_b^{(\rm en)} = N_b^{(\rm lv)} = 
   \,\frac{A}{2} \int_{\tau_i}^{\tau_f} \rho_b(\tau)
   \, {\bar v}(\tau)\, d\tau \, .
 \label{eq15}
 \end{equation}
A similar calculation yields $N_b^{(\rm en)}=N_b^{(\rm lv)}\approx 
N_b^{(i)}{\bar v}_i/2\Delta\eta$. 
$N_b^{(\rm en)}$ and $N_b^{(\rm lv)}$ fluctuate independently;
one therefore expects that the ratio of the mean square fluctuation 
of the number of exchanged baryons $N_b^{(\rm ex)}$ to the average 
initial fluctuation is:
 \begin{equation}
   {(\Delta N_b^{(\rm ex)})^2 \over (\Delta N_b^{(i)})^2} \approx
   {{\bar v}_i \over \Delta\eta} \, , 
 \label{eq16}
 \end{equation}
which is smaller than unity for 
$\Delta\eta{\,\ge\,}{\bar v}_i{\,\approx\,0.65}$.

We conclude that the short time between hadronization and final 
freeze-out precludes the readjustment of net baryon number 
fluctuations in rapidity bins $\Delta\eta\ge 1$. A similar 
calculation applies to net charge fluctuations. Several refinements 
of our estimate are possible but are expected to partially cancel 
each other: Additional transverse expansion lets the temperature drop 
faster than in the Bjorken scenario. During hadronization cooling is 
impeded by the large change in the entropy density between QGP and HG. 
And finally, the short mean free path of baryons in hot hadronic 
matter will significantly reduce our above estimates of the 
dissipation of an initial state fluctuation.

In conclusion, we have argued that the difference in magnitude of 
local fluctuations of the net baryon number and net electric charge 
between confined and deconfined hadronic matter is partially frozen
at an early stage in relativistic heavy-ion collisions. These
fluctuations may thus be useful probes of the temporary formation of 
a deconfined state in such collisions. The event-by-event 
fluctuations of the two suggested observables for collisions with a 
fixed value of the transverse energy $dE_{\rm T}/dy$ or of the 
energy measured in a zero-degree calorimeter would be appropriate
observables that could test our predictions. Further discrimination 
can be achieved by measuring the beam energy dependence of the 
fluctuations: In the QGP the ratio $(\Delta Q/\Delta N_b)^2={5\over 2}$ 
of charge to baryon number fluctuations is a beam-energy independent 
constant; in the HG phase it shows a significant beam energy 
dependence between SPS and RHIC/LHC energies.

This work was supported by the U.S. Department of Energy,
the Japanese Ministry of Education, Science, and Culture,
and the Japanese Society for the Promotion of Science. B.M. thanks 
H.~Minakata (Tokyo Metropolitan University) for his hospitality. 
We are indebted to K.~Rajagopal for valuable discussions. 

{\it Note added:} After finishing this work we received a paper by 
Jeon and Koch \cite{JK00} who discuss similar issues. At the SPS they 
get $(\Delta Q)^2/S\big\vert_{\rm HG} \approx 0.13$ which is more than 
twice our value, due to a smaller resonance decay correction to 
$(\Delta Q)^2$ (30\% instead of our 50\%) and their omission of a 35\% 
extra contribution to $S$ \cite{SH92} from heavy particles (mostly the 
net baryons and strange hadrons).


\begin{references}
\bibitem{fn1}
  Our formulae ignore strange particles. This will do for a 
  semiquantitative understanding. Strange particles carry an 
  additional chemical potential $\mu_{\rm S}$ to ensure 
  strangeness conservation. It vanishes as $\mu\to 0$. In Pb-Pb 
  collisions at the SPS, $\mu_{\rm S}/T{\,\approx\,}0.42$ at 
  hadrochemical freeze-out while $\mu/T{\,\approx\,}1.58$ 
  \cite{PBM99}. Our numerical examples below take this into account.
\bibitem{fn2}
  We emphasize that measuring $(\Delta N_b)^2$ requires the detection 
  of (anti)neutrons. The net proton number is not conserved, and its
  fluctuations are dominated by isospin fluctuations induced by the 
  fluctuating $\pi^+/\pi^-$ ratio \cite{JK99}.
\end{references}
\end{document}